\documentclass[10pt, twocolumn]{article}

\usepackage[T1]{fontenc}
\usepackage[utf8]{inputenc}

\usepackage{geometry}
\usepackage{amsmath}
\usepackage{graphicx}
\usepackage{microtype}
\usepackage{sectsty}
\usepackage{authblk}
\usepackage{caption}
\usepackage[style]{abstract}
\usepackage{cite}
\usepackage{mcite}
\usepackage{upgreek}

\sectionfont{\normalsize}
\renewcommand{\abstitlestyle}[1]{}

\title{Light-nuclei gluons from dijet production in proton--oxygen collisions}

\author{Petja Paakkinen\footnote{email: petja.k.m.paakkinen@jyu.fi}}

\affil{\small
  Instituto Galego de F{\'\i}sica de Altas Enerx{\'\i}as (IGFAE), Universidade de Santiago de Compostela, E-15782 Galicia-Spain
\\
  University of Jyväskylä, Department of Physics, P.O. Box 35, FI-40014 University of Jyväskylä, Finland\footnote{current address}
\\
  Helsinki Institute of Physics, P.O. Box 64, FI-00014 University of Helsinki, Finland
}

\date{\vspace{-0.2em}\large November 9, 2021}

\begin{document}

\twocolumn[
  \maketitle
  \vspace{-1.2em}
  \begin{onecolabstract}
    The prospects of measuring a single-differential dijet cross section during the proposed short proton--oxygen data taking at LHC Run~3 are studied using next-to-leading-order perturbative QCD predictions. With reasonable experimental cuts and luminosity estimates, such a measurement is found to be feasible, and a few inverse nanobarns of integrated luminosity is estimated to be enough to yield new constraints on parton distribution functions (PDFs) of light nuclei. In the absence of a dedicated proton--proton reference, a mixed-energy nuclear modification ratio is proposed for cancelling free-proton PDF uncertainties to obtain a direct access to the nuclear modifications of the parton distributions in oxygen.
  \end{onecolabstract}
  \vspace{1.0em}
]
\saythanks

\section{Introduction}

During its very successful Runs 1 and 2, the Large Hadron Collider (LHC) has contributed significantly to the understanding of the partonic structure of protons~\cite{NNPDF:2017mvq,Hou:2019efy,Bailey:2020ooq} and heavy nuclei alike~\cite{Eskola:2016oht,AbdulKhalek:2020yuc,Kusina:2020lyz,Duwentaster:2021ioo}. The latter, described in terms of nuclear parton distribution functions (nPDFs), has benefited particularly from the precise measurements of D$^0$-meson and dijet production in 5.02 TeV proton--lead~(pPb) collisions~\cite{Aaij:2017gcy,Sirunyan:2018qel}, giving strong evidence of nuclear gluon shadowing and antishadowing phenomena~\cite{Kusina:2017gkz,Eskola:2019dui,Eskola:2019bgf}. The main shortcoming of these new constraints for the nPDF global analyses is, however, that they probe the gluon content only at the very heavy end of the nuclear mass-number spectrum. If one then tries to deduce the gluon distributions of the lighter nuclei from these measurements, they will inevitably run into problems with a strong parametrisation dependence.

Some aid for constraining the nuclear-mass dependence of the gluon PDF can be expected from charm production at the LHCb experiment in the fixed-target mode~\cite{LHCb:2018jry}, but only the high-$x$ region of the nPDFs can be accessed through these measurements, and the mass-number systematics of the nuclear shadowing~\cite{Armesto:2006ph}, or the anticipated onset of gluon saturation~\cite{Morreale:2021pnn}, need to be extracted from elsewhere. Also, even though the capabilities of accelerating different nuclei at the Relativistic Heavy-Ion Collider (RHIC) have been demonstrated, no plans for proton--light-ion
data taking beyond the 2015 proton--aluminium run have been included in the beam-use proposals for the remainder of the collider lifetime.

With the above limitations on the available data and future prospects in mind, the proposed short proton--oxygen~(pO) data-taking at the LHC Run~3~\cite{Citron:2018lsq,Brewer:2021kiv} with 9.9 or 9.0 TeV collision energy could provide a unique short-term opportunity for studying the light-nuclei gluons before the advent of the Electron-Ion Collider~\cite{Accardi:2012qut,AbdulKhalek:2021gbh,Aschenauer:2017oxs}. Compared to the more extensive pPb runs, there are two competing factors that affect the expected statistics: on one hand, the smaller charge of the fully stripped oxygen ions makes it possible to have more ions per bunch, increasing the instantaneous luminosity, but on the other hand, the limited run time restricts the achievable integrated luminosity. Estimates for the attainable values have ranged from $0.2\ {\rm nb}^{-1}$~\cite{Citron:2018lsq} up to approximately $6\ {\rm nb}^{-1}$ delivered to the CMS and ALICE experiments~\cite{Jebramcik:2019eot,Brewer:2021kiv}. Similarly, while the pO cross section is smaller than the pPb one approximately by a factor $16/208$ at a fixed collision energy, the increase in the latter by almost a factor of two compared to the aforementioned measurements in the $5.02$ TeV pPb collisions compensates for this loss.

In this Letter, the prospects of performing a single-differential dijet measurement within the short LHC pO run are studied with theory predictions performed at next-to-leading order (NLO) perturbative QCD. The expected statistics with the luminosity estimates above are evaluated, and the possible impact on the nPDFs is discussed.
Theoretical uncertainties from free-proton PDFs are also quantified and possibilities to reduce them with different ratios in the absence of same-energy proton--proton~(pp) reference are assessed.

\section{Nuclear-mass dependence of gluon PDF}

\begin{figure}[htb]
  \centering
  \includegraphics[width=\columnwidth]{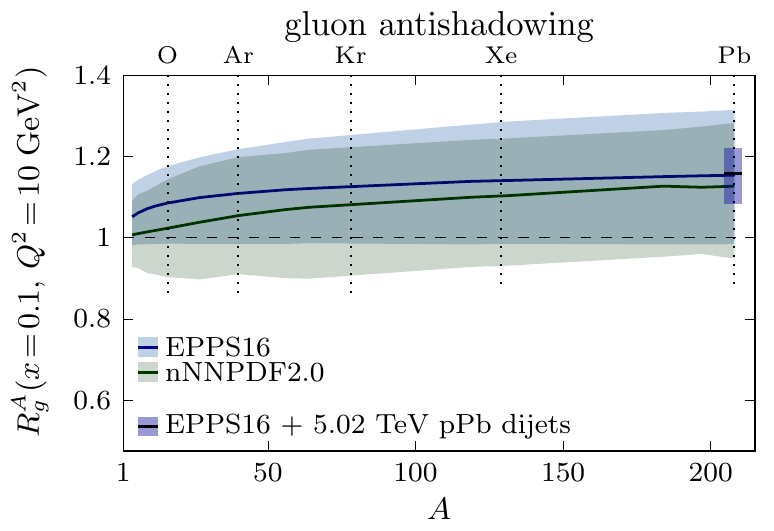}
  \includegraphics[width=\columnwidth]{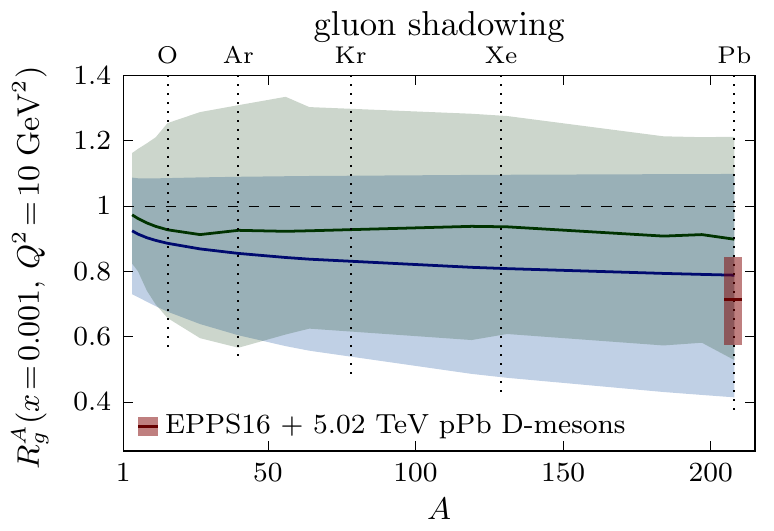}
  \caption{The nuclear modification factors for gluon PDF in the antishadowing (top) and shadowing (bottom) regions as a function of the nuclear mass number from the EPPS16~\protect\cite{Eskola:2016oht} and nNNPDF2.0~\protect\cite{AbdulKhalek:2020yuc} analyses. Results from Hessian PDF reweighting studies~\protect\cite{Eskola:2019dui,Eskola:2019bgf} are also indicated.}
  \label{fig:gluon_A_dep}
\end{figure}

\begin{figure}[htb]
  \centering
  \includegraphics[width=\columnwidth]{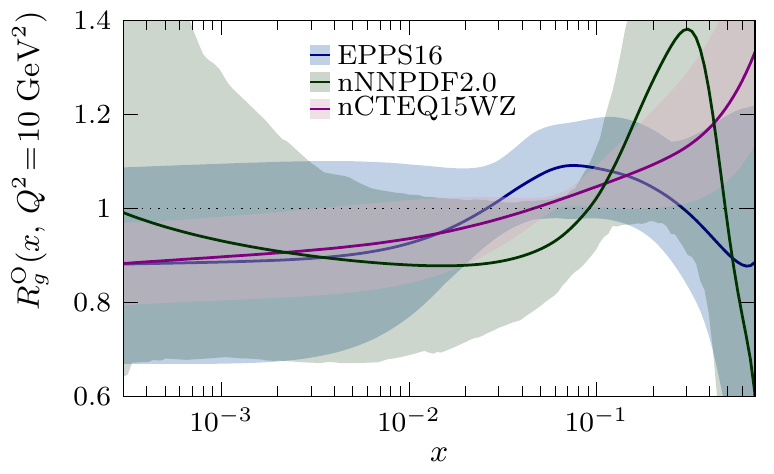}
  \caption{The gluon nuclear modification factor for oxygen as a function of the momentum fraction $x$ from the EPPS16~\protect\cite{Eskola:2016oht}, nNNPDF2.0~\protect\cite{AbdulKhalek:2020yuc} and nCTEQ15WZ~\protect\cite{Kusina:2020lyz} nPDFs.}
  \label{fig:gluon_O}
\end{figure}

The mass-number dependence of gluon nuclear modification factor
\begin{equation}
  R_g^A(x,Q^2) = \frac{f_g^A(x,Q^2)}{A f_g^p(x,Q^2)},
\end{equation}
where $f_g^A$ is the gluon PDF of a nucleus with mass number $A$ and $f_g^p$ the gluon PDF of the free proton, is shown in Figure~\ref{fig:gluon_A_dep} at the scale $Q^2 = 10~{\rm GeV}^2$ for momentum fractions $x = 0.1$ (antishadowing region) and $x = 0.001$ (shadowing region) for the EPPS16~\cite{Eskola:2016oht} and nNNPDF2.0~\cite{AbdulKhalek:2020yuc} nPDFs. All PDF uncertainties are presented at a 90\% confidence level, in the case of nNNPDF2.0 using the prescription laid out in Ref.~\cite{AbdulKhalek:2019mzd} (Eq.~5.3), and for EPPS16 (and later also for nCTEQ15WZ~\cite{Kusina:2020lyz}) with the conventional asymmetric prescription given in Ref.~\cite{Eskola:2016oht} (Eq.~53). As is evident from the figure, the gluon-PDF uncertainties in these analyses are large throughout the nuclear spectrum.

The nPDFs can be constrained further by using the 5.02 TeV pPb dijet and D$^0$ measurements~\cite{Aaij:2017gcy,Sirunyan:2018qel}, with the results from reweighting the EPPS16 nPDFs with these data~\cite{Eskola:2019dui,Eskola:2019bgf} shown also in the figure. However, any impact on the gluon PDFs of lighter nuclei inferred from these data will heavily depend on the assumed functional $A$ dependence, and for example at $x = 0.1$ nNNPDF2.0 suggests the nuclear effects to die off towards smaller $A$ faster than in EPPS16. Even if one uses physical arguments such that nuclear modifications should be smaller for lighter nuclei, there is still enough functional freedom that the nPDF uncertainties e.g.\ at around the mass of oxygen cannot be expected to be reliably reduced by the measurements at Pb. New measurements with lighter nuclei are therefore indispensable to constrain the gluon PDF mass-number dependence, as has been previously discussed also in Refs.~\cite{Citron:2018lsq,Paukkunen:2018kmm,Helenius:2019lop}.

Figure~\ref{fig:gluon_O} shows the nuclear modification of gluon PDF in oxygen as given by the EPPS16, nNNPDF2.0 and nCTEQ15WZ nPDFs. While all of these analyses include some amount of data constraints for the gluon PDFs of heavy nuclei (mostly from pPb collisions at the LHC), the different assumptions on the $A$ dependence lead to very different shapes of modifications in oxygen. Interestingly, they differ significantly in the region $10^{-2} < x < 10^{-1}$, which is highly relevant for studying the parton energy loss in oxygen--oxygen collisions~\cite{Huss:2020dwe,Zakharov:2021uza,Brewer:2021tyv}. Finding direct data constraints for the oxygen gluon PDF would therefore be most timely in order to benefit maximally from the proposed oxygen--oxygen run at the LHC. Extracting the gluon content of oxygen would help also in estimating the nPDF effects in other intermediate-mass nucleus--nucleus collision systems (species used or under consideration at the LHC~\cite{Jebramcik:2019eot} are indicated in Figure~\ref{fig:gluon_A_dep}) before respective proton--nucleus measurements and their implementation in the nPDF analyses.

\section{Dijet production in pO at 9.9 TeV}

The pseudorapidity-differential dijet cross section in proton--nucleus collisions has been demonstrated to be an excellent probe of the gluon nPDF $x$ dependence~\cite{Eskola:2013aya,Eskola:2019dui}. It is therefore interesting to study whether such a measurement could be performed with the short LHC pO run, as will be assessed next. The kinematical cuts are taken to be the same as in Ref.~\cite{Sirunyan:2018qel}, with jets defined through the anti-$k_{\rm T}$ algorithm~\cite{Cacciari:2008gp} with a distance parameter $R = 0.3$. The dijet system is taken to be that composed of the jet with the largest transverse momentum $p_{\rm T}^{\rm lead}$ and the one with the second-to-largest transverse momentum $p_{\rm T}^{\rm sub}$. These are required to satisfy $p_{\rm T}^{\rm ave} = (p_{\rm T}^{\rm lead} + p_{\rm T}^{\rm sub}) / 2 > 55\ {\rm GeV}$ and $p_{\rm T}^{\rm sub} > 20\ {\rm GeV}$, with both residing in $-3 < \eta_\text{jet} < 3$ in the laboratory frame, and to be separated in the azimuthal angle by at least $\Delta\phi > 2\uppi/3$.

The perturbative QCD predictions are performed here at NLO through NLOJet++~\cite{Nagy:2003tz}, with the renormalisation and factorisation scales set equal to $p_{\rm T}^{\rm ave}$. The EPPS16 central prediction for the integrated fiducial NLO parton-level cross section (with the cuts defined above) at 9.9 TeV is
\begin{equation}
  \sigma^{\rm pO}_{\rm 9.9\,TeV} = 81\ {\rm \upmu b}.
\end{equation}
With the $0.2\ {\rm nb}^{-1}$ and $6\ {\rm nb}^{-1}$ luminosities, the expected number of events would therefore be 16000 and 486000, respectively. Assuming a similar better-than-99\% efficiency as has been obtained in the previous measurements in pp and pPb~\cite{CMS:2016kjd}, these values should enable a single-differential measurement, and the higher limit might even allow for placing more stringent cuts on the minimum jet $p_{\rm T}$. Note, however, that the relatively small $R = 0.3$, which in accordance with Ref.~\cite{Sirunyan:2018qel} was chosen to minimise the contribution from the underlying event, makes this observable susceptible to perturbative out-of-cone radiation and non-perturbative hadronisation corrections~\cite{Dasgupta:2007wa,Dasgupta:2014yra,Dasgupta:2016bnd}, whereby the observed hadron-level cross section can be somewhat smaller than the NLO parton-level estimate given above. It would therefore be advisable to aim towards the higher values of luminosity to ensure sufficient statistics and to study whether the cone size could be increased for the pO measurement at these collision energies without inflating the underlying-event contribution. Using the lower 9.0 TeV collision energy would also render the cross section smaller, but a single-differential measurement should still be feasible.

\begin{figure}[htb]
  \centering
  \includegraphics[width=\columnwidth]{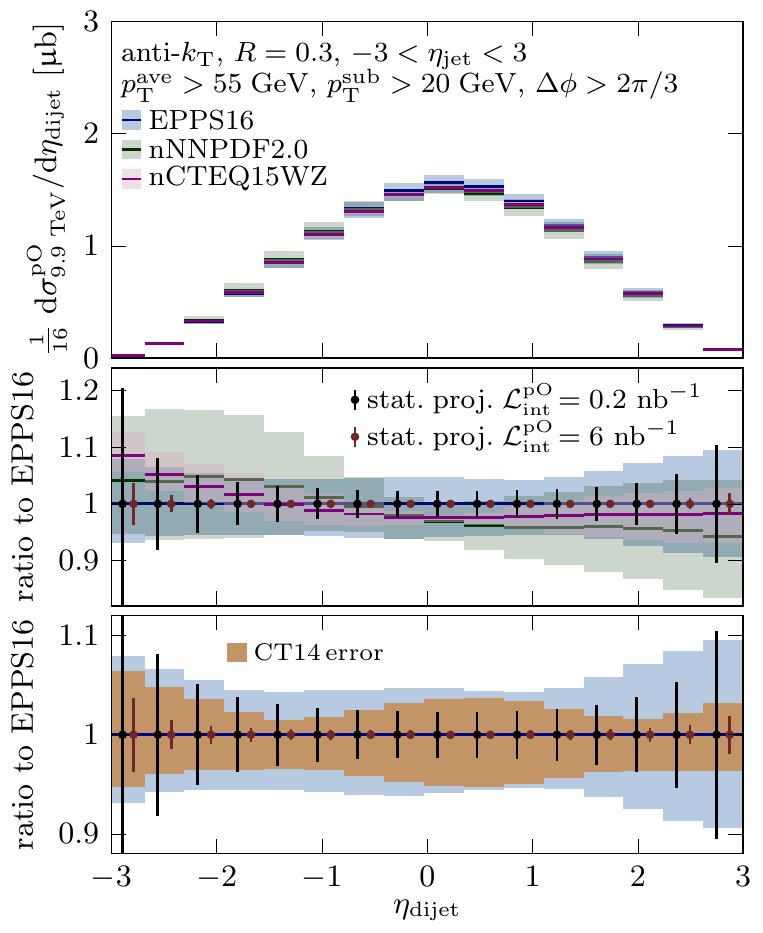}
  \caption{Per-nucleon pseudorapidity-differential parton-level dijet cross section in pO collisions at 9.9 TeV evaluated with the EPPS16~\protect\cite{Eskola:2016oht}, nNNPDF2.0~\protect\cite{AbdulKhalek:2020yuc} and nCTEQ15WZ~\protect\cite{Kusina:2020lyz} nPDFs. Middle and bottom panels show the ratio to EPPS16 central prediction, with the projected statistical uncertainties and baseline free-proton PDF (CT14 NLO~\protect\cite{Dulat:2015mca}) uncertainties indicated.}
  \label{fig:pO_dijet}
\end{figure}

Figure~\ref{fig:pO_dijet} (top panel) shows the predicted per-nucleon single-differential parton-level dijet cross section as a function of the laboratory-frame pseudorapidity of the dijet, defined as
\begin{equation}
	\eta_{\rm dijet} = \frac{1}{2}(\eta^{\rm lead} + \eta^{\rm sub}),
\end{equation}
evaluated with the nPDFs from the EPPS16, nNNPDF2.0 and nCTEQ15WZ analyses. To ease the comparison between the nPDF analyses, the middle panel shows the ratio of the different predictions to the central result from EPPS16. A pattern analogous to that in Figure~\ref{fig:gluon_O} is observed, where at negative rapidities, probing large values of the nuclear $x$, the predictions from nNNPDF2.0 and nCTEQ15WZ are generally above that from EPPS16, whereas at positive rapidities, probing small values of the nuclear $x$, the trend is the opposite.

The projected statistical uncertainties with the luminosity estimates $0.2\ {\rm nb}^{-1}$ and $6\ {\rm nb}^{-1}$ are also shown in Figure~\ref{fig:pO_dijet}, calculated from the expected number of events in each pseudo-rapidity bin. These are generally smaller than the envelope of the predictions, indicating a potential good constraining power. Still, even though the projected statistical uncertainties are found to be smaller than the spread in nPDF predictions already at $0.2\ {\rm nb}^{-1}$, after one accounts for the hadronisation corrections and systematical uncertainties, it can be expected that the data fluctuations would be too large to give strong preference to any particular nPDF set. Therefore, a luminosity of the order of a few inverse nanobarns is expected to be needed to give significant constraints.

A further complication arises from the fact that in proton--nucleus collisions, one is always probing a convolution of proton and nuclear structures. For full consistency with the respective global analyses, the cross sections in this Letter are evaluated for each of the nPDFs with the same free-proton PDFs that were used in the fits. In the case of nCTEQ15WZ the free-proton PDF error sets are not available and therefore only the uncertainties from nuclear degrees of freedom are presented, but for EPPS16 and nNNPDF2.0 the free-proton uncertainties are included. These can be sizeable, as shown in the bottom panel of Figure~\ref{fig:pO_dijet}, where the contribution from the CT14 NLO free-proton PDFs~\cite{Dulat:2015mca} on the EPPS16 uncertainty is presented. By using absolute cross sections, it is therefore very difficult to disentangle nuclear modification effects from the free-proton degrees of freedom, which complicates the interpretation of the measurement and makes the extracted nuclear modifications strongly dependent on the used free-proton baseline PDFs.

\section{Forward-to-backward ratio}

\begin{figure}[htb]
  \centering
  \includegraphics[width=\columnwidth]{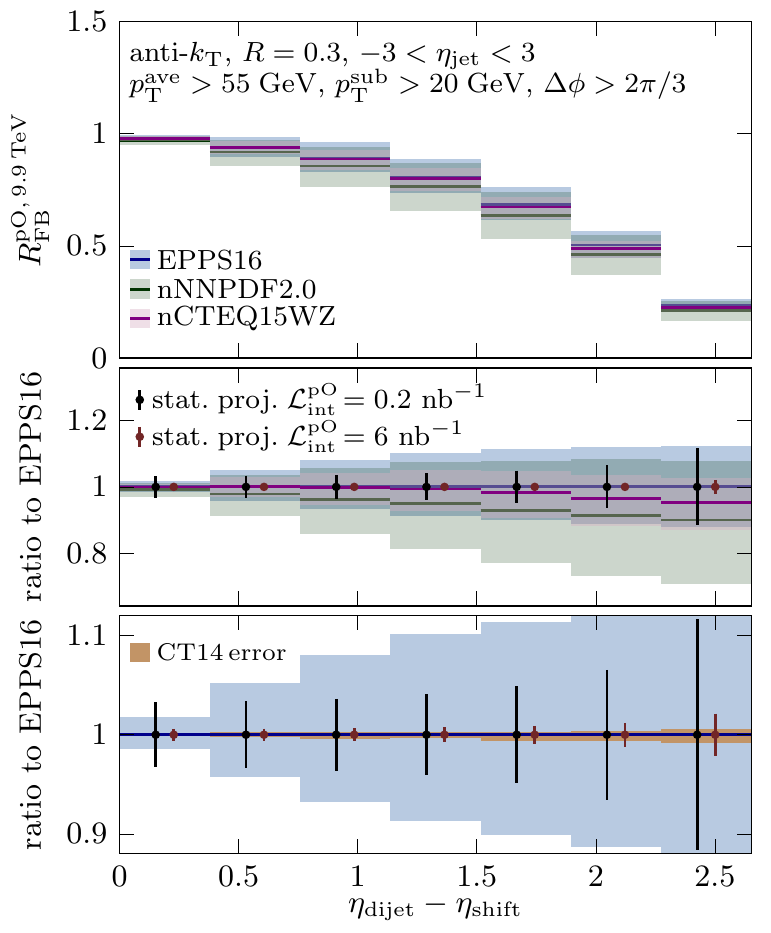}
  \caption{As Figure~\ref{fig:pO_dijet}, but for the forward-to-backward ratio.}
  \label{fig:pO_dijet_FB}
\end{figure}

A typical observable used to reduce free-proton and scale uncertainties as well as experimental systematic uncertainties in proton--nucleus collisions
is the forward-to-backward ratio, discussed in the context of dijet production in Ref.~\cite{Eskola:2013aya}, where one divides the cross sections at positive center-of-mass-frame rapidities with the respective values at negative rapidities. Due to the equal-rigidity acceleration at the LHC, the pO center-of-mass is shifted from the laboratory frame by
\begin{equation}
  \eta_{\rm shift} = \frac{1}{2} \log\frac{16}{8} = 0.347,
\end{equation}
a value common for all isoscalar nuclei. Hence, in the laboratory frame, the forward-to-backward ratio is defined as
\begin{equation}
  R_{\rm FB}^{\rm pO,\,9.9\,TeV}(\eta_{\rm dijet}) = \frac{{\rm d}\sigma^{\rm pO}_{\rm 9.9\,TeV} / {\rm d}\eta_{\rm dijet}(\eta_{\rm dijet})}{{\rm d}\sigma^{\rm pO}_{\rm 9.9\,TeV} / {\rm d}\eta_{\rm dijet}(2\eta_{\rm shift} - \eta_{\rm dijet})}.
\end{equation}
As shown in Figure~\ref{fig:pO_dijet_FB} (bottom panel), this ratio leads to an excellent cancellation of the free-proton-PDF uncertainties, which now remain smaller than the projected statistical uncertainties, and thus gives a good handle on the nuclear modifications themselves.

The problem with this type of observable, however, is that by taking the ratio one loses the locality in the momentum fraction, whereby one is probing only the correlation between high- and low-$x$ nuclear modifications. Thus, even rather different shapes in $R^{\rm O}_g$ can lead to a very similar shape in the forward-to-backward ratio, as can be seen from the top and middle panels of Figure~\ref{fig:pO_dijet_FB}. Moreover, at $\eta_{\rm dijet} = \eta_{\rm shift}$, probing $x$ of values around $0.01$, the ratio goes to unity by construction, and any information on the nuclear modifications in this region is lost.

\section{Mixed-energy nuclear-modification ratio}

A more direct access to the $x$ dependence of the nuclear modifications would be obtained by using the nuclear modification ratio with respect to a pp baseline at the same collision energy. However, the needed pp reference run at 9.9 (or 9.0) TeV is currently not expected to take place during Run~3. Previous workarounds to this problem have included using interpolated or extrapolated pp reference, but this can lead to sizeable parametrisation uncertainty, and a recent study in the context of oxygen--oxygen collisions found that using three reference energies (from the same run to cancel systematical uncertainties) was necessary to construct a precise baseline~\cite{Brewer:2021tyv}.

\begin{figure*}[htb]
  \centering
  \includegraphics[width=\columnwidth]{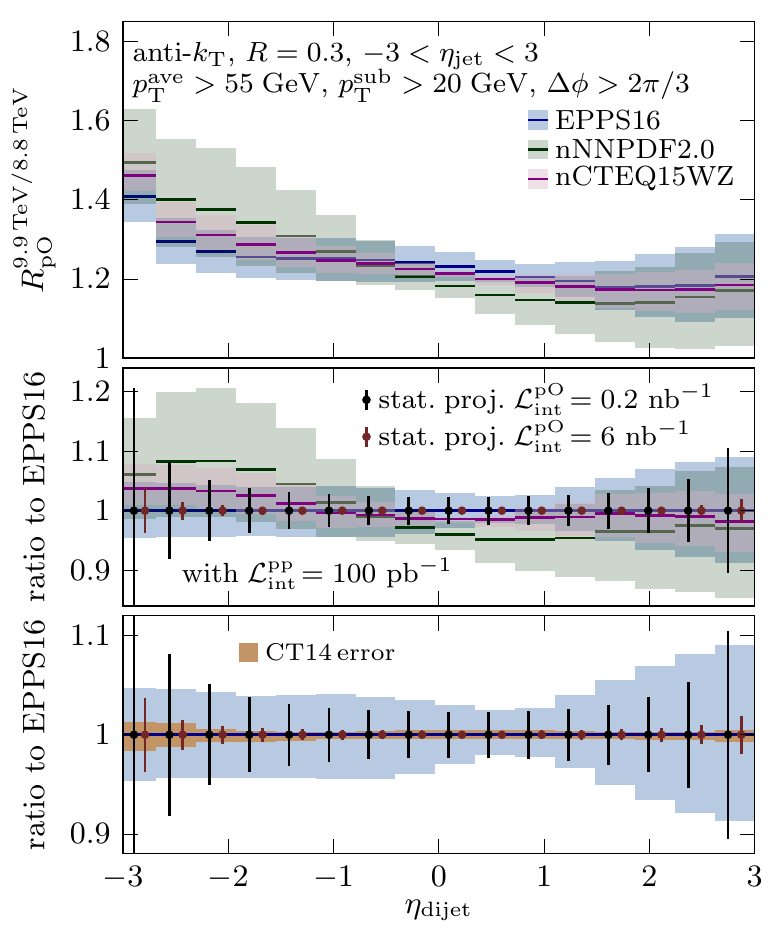}
  \hfill
  \includegraphics[width=\columnwidth]{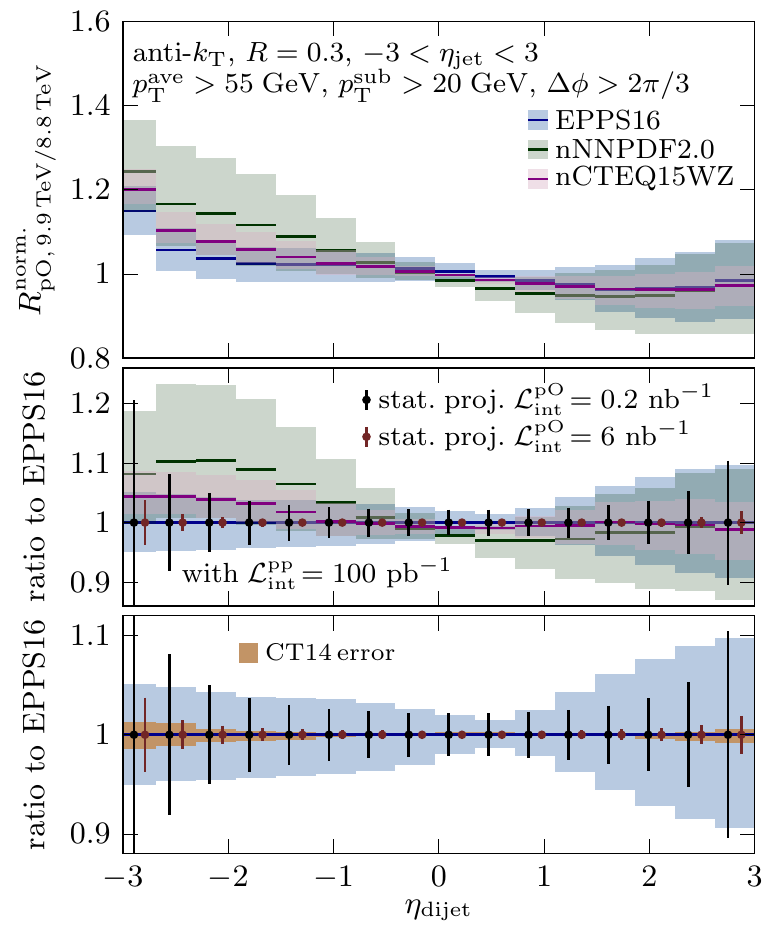}
  \caption{As Figure~\ref{fig:pO_dijet}, but for the mixed-energy nuclear-modification ratio (left), and for the self-normalised double ratio (right).}
  \label{fig:pO_dijet_RpO_normalized_mixed_energy}
\end{figure*}

Another option, suggested also in Ref.~\cite{Brewer:2021tyv}, is to take a ratio between two different, but close-by, energies. This is a viable opportunity for the pO measurement, as a high-statistics pp reference at 8.8 TeV (or 8.0 TeV, depending on the energy of the anticipated pPb run) with as much as $100\ {\rm pb}^{-1}$ could be expected to be taken during Run~3~\cite{Citron:2018lsq}. This mixed-energy nuclear-modification ratio is presented in Figure~\ref{fig:pO_dijet_RpO_normalized_mixed_energy} (top left). Again, the pO rapidity shift needs to be accounted for, and therefore the jets in pp are required to be within the interval $-3.347 < \eta_{\rm jet} < 2.653$ in the laboratory frame and the ratio is defined in terms of a shifted pp reference as
\begin{equation}
  R_{\rm pO}^{\rm 9.9\,TeV/8.8\,TeV}(\eta_{\rm dijet}) = \frac{\frac{1}{16}\,{\rm d}\sigma^{\rm pO}_{\rm 9.9\,TeV} / {\rm d}\eta_{\rm dijet}(\eta_{\rm dijet})}{{\rm d}\sigma^{\rm pp}_{\rm 8.8\,TeV} / {\rm d}\eta_{\rm dijet}(\eta_{\rm dijet} - \eta_{\rm shift})}.
\end{equation}

Interpreting the ratio is not as straightforward as in the same-energy case. Due to the steeply growing nature of gluon PDFs at small $x$, even a small shift in the probed value can cause a significant change in the cross section. For this reason, there is a 20\% enhancement at midrapidity simply from using a lower-energy pp reference, and the effect grows to almost 50\% at $\eta_{\rm dijet} = -3$. The free-proton uncertainties are still well under control, as can be seen from the bottom left panel of Figure~\ref{fig:pO_dijet_RpO_normalized_mixed_energy}. This follows since at large perturbative scales, the PDFs at two close-by values of $x$ are strongly correlated through the DGLAP evolution (but also due to the way they are parametrised). Since the nPDF analyses are in any case moving towards accounting for the full correlations with free-proton PDFs~\cite{AbdulKhalek:2019mzd,AbdulKhalek:2020yuc,Eskola:2021mjl}, the remaining small free-proton uncertainty is completely acceptable.

Due to the good cancellation of free-proton PDF uncertainties in the $R_{\rm pO}^{\rm 9.9\,TeV/8.8\,TeV}$ ratio, the constraining power on the nuclear modifications is significantly improved. In particular, there is now less overlap in the predictions from EPPS16 and nNNPDF2.0 compared to the absolute cross sections in Figure~\ref{fig:pO_dijet}, showing that this observable is able to resolve different nPDF parametrisations. Again, as shown in Figure~\ref{fig:pO_dijet_RpO_normalized_mixed_energy} (middle left panel), the projected statistical uncertainties with $0.2\ {\rm nb}^{-1}$ are smaller than the spread in the predictions, but to ensure good constraining power after all experimental uncertainties are accounted for, an integrated luminosity in the few-inverse-nanobarns range would be preferred. It should be noted also that using the same pp reference for the nuclear modification ratios of pO and pPb at Run~3 makes these measurements correlated, and for a reliable extraction of the nPDFs, it would be optimal to publish these cross correlations as well.

One could also consider using the self-normalised ratio
\begin{multline}
  R_{\rm pO,\,9.9\,TeV/8.8\,TeV}^{\rm norm.}(\eta_{\rm dijet}) \\= \frac{\frac{1}{\sigma^{\rm pO}_{\rm 9.9\,TeV}} {\rm d}\sigma^{\rm pO}_{\rm 9.9\,TeV} / {\rm d}\eta_{\rm dijet}(\eta_{\rm dijet})}{\frac{1}{\sigma^{\rm pp}_{\rm 8.8\,TeV}} {\rm d}\sigma^{\rm pp}_{\rm 8.8\,TeV} / {\rm d}\eta_{\rm dijet}(\eta_{\rm dijet} - \eta_{\rm shift})},
\end{multline}
shown in the top right panel of Figure~\ref{fig:pO_dijet_RpO_normalized_mixed_energy}, as in Refs.~\cite{Sirunyan:2018qel} and~\cite{Eskola:2019bgf}. The advantage with this double ratio is that the luminosity and hadronisation uncertainties cancel separately for both pO and pp. This ratio also leads to an improved reduction of the free-proton PDF uncertainties at midrapidity (see the bottom right panel), but with the expense that also part of the nuclear modification uncertainties cancel, diminishing the potential constraining power.

While it would be possible to study the expected impact on the nPDFs in more detail by using the reweighting methods~\cite{Giele:1998gw,Ball:2010gb,Ball:2011gg,Watt:2012tq,Sato:2013ika,Paukkunen:2013grz,Paukkunen:2014zia,Eskola:2019dui,Schmidt:2018hvu,Hou:2019gfw}, this is not pursued here for the following reasons: first, doing so reliably would require estimating the systematical uncertainties, which is outside the scope of this Letter, and second, as the nuclear modifications from different nPDFs are barely overlapping in some places, there appears to be a strong parametrisation dependence in them, and it is not guaranteed that such a study would reflect the true impact in a full analysis where some of the parametrisation assumptions could be relaxed.

\section{Conclusion}

Summarising, it has been shown in this Letter that measuring dijet production in the 9.9 (or 9.0) TeV pO collisions during the LHC Run~3 would significantly help in understanding the nuclear-mass-number dependence of the gluon PDF, and as a rough estimate, an integrated luminosity of the order of a few inverse nanobarns should be enough to perform a single-differential measurement with meaningful constraints on the nPDFs. Moreover, with the expected 8.8 (or 8.0) TeV pp reference run, it would be possible to measure a mixed-energy nuclear modification ratio, providing a direct access to the gluon nuclear modification factor without a strong dependence on the free-proton PDFs.
This ratio was shown to give better resolution on the different nPDF parametrisations than the forward-to-backward ratio, in which one loses part of the information. These results corroborate the usefulness of even a short pO data taking during LHC Run~3, in addition to the motivation from cosmic-ray physics~\cite{Albrecht:2021yla}.

\paragraph{Aknowledgments}

The author would like to thank J.~Brewer, A.~Mazeliauskas, and W.~van~der~Schee for the organisation of the ``Opportunities of OO and pO collisions at the LHC'' workshop, for which the first results of this study were prepared, and for many fruitful discussions; C.~Andrés, K.~J.~Eskola, H.~Paukkunen and C.~A.~Salgado also for useful discussions; and F.~Olness for providing the nCTEQ15WZ nPDF grid files. This work has received financial support from Xunta de Galicia (Centro singular de investigación de Galicia accreditation 2019-2022), by European Union ERDF, and by the “María de Maeztu” Units of Excellence program MDM-2016-0692 and the Spanish Research State Agency, from European Research Council project ERC-2018-ADG-835105 YoctoLHC, and from the Academy of Finland, project nr.~330448.

\bibliographystyle{phaip_mcite}
\bibliography{pO_paper}

\end{document}